# Role of non-timber forest products in sustaining forest-based livelihoods and rural households' resilience capacity in and around protected area: a Bangladesh study

Sharif Ahmed Mukul[1,2,3]*, A. Z. M. Manzoor Rashid[4], Mohammad Belal Uddin[4], Niaz Ahmed Khan[5]

[1]*Tropical Forestry Group, School of Agriculture and Food Sciences, Faculty of Science, The University of Queensland, Brisbane, QLD 4072, Australia*

[1]*School of Geography, Planning and Environmental Management, Faculty of Science, The University of Queensland, Brisbane, QLD 4072, Australia*

[3]*Centre for Research on Land-use Sustainability, Maijdi, Noakhali 3800, Bangladesh*

[4]*Department of Forestry and Environmental Science, School of Agriculture and Mineral Sciences, Shahjalal University of Science and Technology, Sylhet 3114, Bangladesh.*

[5]*Department of Development Studies, University of Dhaka, Dhaka 1000, Bangladesh*

*Corresponding author: sharif_a_mukul@yahoo.com / s.mukul@uq.edu.au

**Abstract**

People in developing world derive a significant part of their livelihoods from various forest products, particularly non-timber forest products (NTFPs). This article attempts to explore the contribution of NTFPs in sustaining forest-based rural livelihood in and around a protected area (PA) of Bangladesh, and their potential role in enhancing households' resilience capacity. Based on empirical investigation our study revealed that, local communities gather a substantial amount of NTFPs from national park despite the official restrictions. 27% households (HHs) of the area received at least some cash benefit from the collection, processing and selling of NTFPs, and NTFPs contribute as HHs primary, supplementary and emergency sources of income. NTFPs also constituted an estimated 19% of HHs net annual income, and were the primary occupation for about 18% of the HHs. HHs dependency on nearby forests for various NTFPs varied vis-à-vis their socio-economic condition as well as with their location from the park. Based on our case study the article also offers some clues for improving the situation in PA.

## Introduction

Forests and forest products have played a vital role in sustaining the livelihoods of poor, forest-dependent communities for centuries. According to the World Bank (2002), more than 1.6 billion people throughout the world rely on forests for their livelihoods, and some 350 million people depend solely on forests, both for their subsistence and incomes. Another study suggests that over two billion people, a third of the world's population, use biomass fuels, mainly firewood, to cook and heat their homes and rely on traditional medicines harvested from forests (Arnold et al. 2006). The same study also finds that, in some 60 developing countries, hunting and fishing on forested land supplies over a fifth of people's protein requirements (Mery et al. 2005). During the past two decades, there has been an increased recognition of the significant role of various forest products for household food and livelihood security, with an appreciation of the major role that NTFPs play.[1] In fact, for a large number of people, NTFPs are still a more important resource than timber. Wunder (2000) estimated that smallholder families living in forest margins in diverse parts of the world earn between 10% and 25% of their household income from NTFPs. Another study suggests that tropical forests in parts of southeast Asia provide as much as $50 per month per hectare to local people and that this comes only from exploiting NTFPs (Sedjo 2000). In recent times, NTFPs have also gained national and international attention due their perceived conservation potential; there is a belief that the collection and use of NTFPs are ecologically less destructive than timber harvesting (Anderson 1990; Plotkin and Famolare 1992; Arnold and Ruiz Pérez 2001; Gubbi and MacMillan 2008; Mukul et al. 2010).

Asia is undoubtedly the world's largest producer and consumer of NTFPs (Vantomme et al. 2002); a study by de Beer and McDermott (1996) showed that about 27 million people in South East Asia rely on NTFPs. This situation is no different in Bangladesh, where the collection, processing, and selling of NTFPs provide major employment opportunities for the poorest rural population of nearly 300,000 people with an annual contribution of about Tk1.3[2] billion to the country's economy (Basit 1995; GoB 1993). The operation of small-scale enterprises and the manufacture of secondary products from major NTFPs like bamboos and rattans is also a low-cost but effective strategy for employment generation for semi-skilled laborers in urban-rural fringes in Bangladesh (Alamgir et al. 2005, 2006, 2007; Mukul 2011; Mukul and Rana 2013).

Alongside a renewed focus on the role of NTFPs in conservation, the past few decades have also witnessed a growing interest and effort in establishing protected areas (PAs) worldwide as a key strategy to address the problem of massive forest and biodiversity loss (Hockings 2003; Watson et al. 2014). Currently, PAs are amongst the world's dominant land uses, covering more area than all the agricultural crops combined (UNEP-WCMC 2014). They are very often argued to be the most effective measure for conserving nature and natural resources, and are considered to be the cornerstone of many national and regional conservation strategies (Lewis 1996; Mulongoy and Chape 2004). In addition to their conservation role, another stated objective of establishing PAs is to promote a certain degree of equity in resource generation and distribution, and to contribute to people's livelihoods, especially the forest-dependent people living within the PA's locality (Gubbi and MacMillan 2008).

---

[1] According to the Food and Agriculture Organization of the United Nations (FAO), NTFPs are products of biological origin other than wood derived from forests, other wooded land, and trees outside forests. Examples include fruits, firewood, bamboo, rattans, medicinal plants, spices, wildlife, and wildlife products.
[2] At the time of this study, US1$ = Tk 69 (69 Bangladesh Taka).

Our knowledge and understanding about the role of NTFPs vis-à-vis the lives and livelihoods of local communities in PAs are still inadequate (Mukul et al. 2010). Unfortunately, in many tropical, developing countries, traditional, forest-dependent people are either denied or are weakly recognized during the formulation and manageent of PAs (seefor example, Gadgil 1990; Gubbi and MacMillan 2008; Mukul and Quazi 2009; Mukul et al. 2008, 2012, 2014). A good number of studies have reported the importance and contribution of NTFPs in local, forest-reliant livelihoods in some tropical developing countries (see, for example, Malhotra et al. 1991; Ganesan1993; Gunatillike et al. 1993; Townson 1995; Cavendish 2000; Malla 2000; Mallik 2000; Ambrose-Oji 2003; Mahapatra et al. 2005). Only a few of these studies, however, focus on the situation in PAs where, ideally, there is greater emphasis on conservation than on livelihoods (Gubbi and MacMillan 2008). Furthermore, due to a rapidly changing global environment and, sometimes due to unanticipated environmental events, such livelihoods are facing greater exposure to risk and vulnerabilities than at any time before (Fisher et al. 2010). There are evidences where local communities depends on forests and associated natural resources to improve their capacity for resilience in such a context (Mertz et al. 2009; Fisher et al. 2010).

Against this backdrop, our study was an attempt to explore the role of NTFPs in the life and livelihood of local people in and around a north-eastern PA in Bangladesh, as well as to understand the significance of NTFPs as a mean of improving households' resilience. As noted earlier, research on the subject is very limited, and this research, in its own modest ways, is expected to contribute to the gap in the literature, and also illuminate the process of better understanding the interconnectivity between NTFPs and local people, particularly in the context of PAs and thereby prompting a rethinking about existing PA management system at the policy planners' level.

**Materials and methods**

*Background of the study site*

To date, 37 PAs have been notified in Bangladesh,[3] under the International Union for Conservation of Nature's (IUCN) protected area management categories II and IV, covering nearly 11% forest area of the country (Dudley (2008). For the present study, we purposively selected Satchari National Park considering its particular richness in biodiversity, convenient access from the capital city Dhaka and nearby Sylhet , and its broad representativeness of the mainstream PAs of the country (Uddin et al. 2013). The park is also one of the PAs in which a co-management approach was pioneered in the country (Mukul et al. 2014).

The word 'Satchari' comes from "seven streams" (locally called *'chara'*) and refers to the streams that flow through the forest. The area of the park is 243 ha which comprises of the forests of Raghunandan Hills Reserve within the Satchari Range (Uddin and Mukul 2007)..  Administratively the park is located in Chunarughat Upazilla (administrative entity; sub-district) of Habiganj district and is situated nearly 130 km north-east of capital city Dhaka.

---

[3] See, for more detail, http://www.bforest.gov.bd/index.php/protected-areas (accessed on 11 December, 2014).

India borders the park on its southern part (Figure 1). Other adjacent areas are covered by tea estates, rubber gardens, agar plantations and paddy fields.

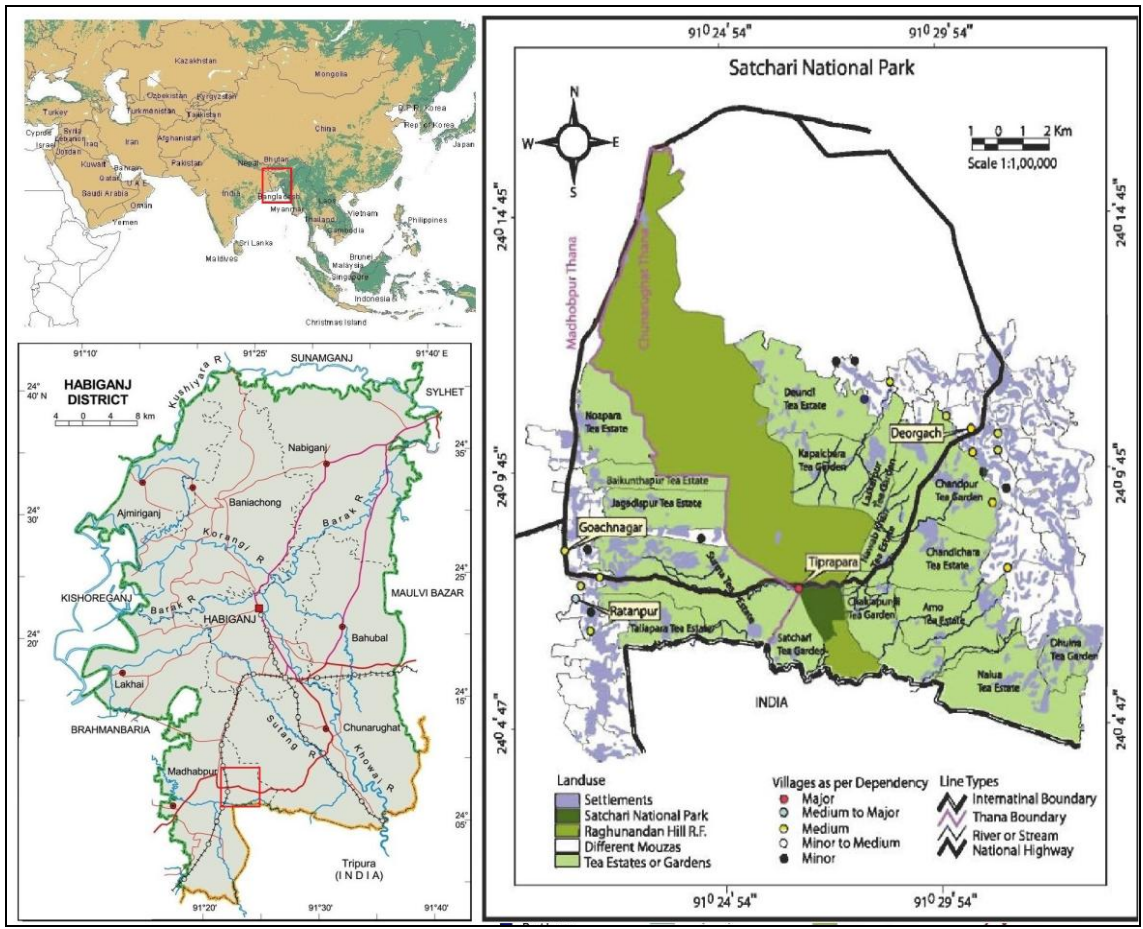

**Figure 1.** Location map of the study site

The vegetation of the park is 'evergreen', however large-scale conversion of the indigenous forest cover to plantations has changed its classic forest type entity (Uddin et al. 2013). Now only 200 ha of land contain natural forests; the rest is secondary raised forest. It is also one of the last habitats of critically endangered Hollock gibbons (*Hoolock hoolock)* in the country as well as in the sub-continent.

The park has an undulating topography with slopes and hillocks ( locally called *tilla*) ranging from 10 to 50m in elevation. A number of small, sandy-bedded streams drain the forest, all of which dry out in the winter season after November. The annual average rainfall is 4162 mm. July is the wettest month having an average of about 1250 mm of rain, while December is the driest with no rainfall. May and October, the hottest months, have an average maximum temperature around $32^0$C, while January is the coldest month when the minimum temperature drops to about $12^0$C. The relative humidity of the study area is about 74% during December while it is over 90% during July-August.

*Sampling protocol*

A total of 19 villages with varied degrees of interests in the national park as well as in the reserved forest (RF) has been identified (Mollah et al. 2004). Of these villages, one village is located within the national park area and is inhabited by an ethnic community, the *Tripura* tribe. The other settlements that have stakes in the national park are located about 3-8 kilometers away from the park. Table 1 lists the degree of dependency that the various villages have on the RF as well as on the national park. Local people have traditionally collected various resources from the national park and adjacent RF. Many households, particularly poor households from the 19 villages studied, rely entirely or partially on the RF as well as on the park for collecting firewood, timber, bamboo, herbal remedies, and other NTFPs.

**Table 1.** Degree of dependency of various villages on the park and RF

| **Degree of dependency** | **Name of the villages** |
| --- | --- |
| Major | *Tiprapara* |
| Medium to major | Gazipur, *Ratanpur* |
| Medium | Kalishiri, Ghanoshyampur, Doulatkhabad, *Deorgach* |
| Minor to medium | Baghbari, Teliapara, *Goachnagar*, Ektiarpur, Marulla, Nayani Bongaon |
| Minor | Shanjanpur, Rasulpur, Promnandapur, Bhaguru, Enatbad, Holholia |

**Source**: Mollah *et al*. (2004); Names of the study villages are Italicized

*Data collection*

We collected both quantitative and qualitative data. We randomly selected one village from each of the first four 'forest dependence categories' as identified by Mollah et al. (2004), including the only village inside the park, Tiprapara. We did not consider any village with only a minor degree of dependence on the PA and adjacent RF. After preliminary field visits and a reconnaissance survey, however, we performed a necessary ranking change between Deorgach and Ratanpur, given the fact that the forest is more critical to the people of Ratanpur than to the people of Deorgach.

We conducted household survey in four sample villages—Tiprapara, Ratanpur, Deorgach, and Goachnagar—within a one-year time span (January to December, 2006). Before household surveys, we arranged some focus group discussions to construct a community map and a community profile of the study villages (Table 2). We also conducted a number of field visits to observe and verify the information recorded during our community mapping exercises. We interviewed 101 households in our four sample villages from a total households of 818 households within the survey villages.

**Table 2.** Profile of the study villages

| Village | Location | Population | Sample |
| --- | --- | --- | --- |

|  | and distance | Size (HHs) | size (n) | Forest practices* |
|---|---|---|---|---|
| Tiprapara | Inside (0 km) | 22 | n = 22 | Collect fuelwood, house building materials, fruits and other NTFPs, cultivate lemon and others |
| Ratanpur | Outside (2 km) | 156 | n = 16 | Mainly involved with illegal tree felling, and collecting fuelwood |
| Deorgach | Outside east (3 km) | 316 | n = 31 | Mainly collect fuelwood, some involved with illegal tree felling |
| Goachnagar | Outside west (4 km) | 328 | n = 32 | As above |

* As described by Mollah et al. (2004)

Several methodological choices are available for assessing households' income and dependency on forests (see Wollenberg and Nawir 1998; Wollenberg 2000; IIED 2003; Vedeld et al. 2004). To measure a household's forest dependency, we considered the annual cash contribution of the forest to households' livelihoods. Accordingly, for convenience of data analysis during the time of the focus group discussions, the average incomes from the forest for three distinct forest dependency categories were estimated respectively as: Tk 54 000 per annum or above for the most forest-dependent category; Tk 54 000 to Tk 24 000 per annum for the moderately forest-dependent category; and below Tk 24 000 per annum for the least forest-dependent category. Additionally, households were categorized into three income classes: extremely poor households (those with monthly incomes below Tk 2000); medium to poor households (those with monthly incomes below Tk 7500 but above Tk 2000); and rich households (those with monthly incomes of Tk 7500 or more). For estimating a household's net income from forest, we used following formula, as used by Ambrose-Oji (2003):

> ***Net Forest Income:*** *Direct cash benefits from selling of all harvested forest products (revenue) + Market value of the consumed forest products which they may have otherwise purchased from the market (savings) – investment cost / opportunity cost.*

In Tiprapara, we took a 100% sample, as villagers were highly dependent on the park. In other villages, we took a 10% sample of households from each of the forest dependency classes using a stratified random sampling approach. During household surveys we used a semi-structured questionnaire. The questionnaire allowed us to record details about the households' NTFPs and other products collection from the forest, quantity harvested, frequency of harvesting, the forest origin and final use (consumed, traded or gifted) of these products, and income from various forest products (i.e. NTFPs, timber andothers). We additionaly collected information on market potential of different NTFPs available in the area and their possible contribution to the households' socio-economic advancement. We performed Students *t*-test and analysis of variance (ANOVA) to compare the means as well as to check any significant difference between the observations.

## Results

*Rural livelihoods in and around Satchari*

The primary occupation observed in the study villages were: agriculture (37%), mainly paddy cultivation, followed by NTFP extraction (18%), timber poaching (18%), day labor (15%), small business (5%), government and non-government services (4%), and overseas employment (2%). The scenario was different in Tiprapara; as there was no land suitable for agricultural practice the villagers were found to work mostly as day labor (38.5%) followed by NTFPs extraction (mainly firewood, some 32%). Forest patrolling was the main service performed by the residents of Tiprapara. In addition, day laborers from all the villages also collected firewood in their off-days and during agricultural off periods. Based on the income classes as described earlier 37% of the HHs were extremely poor followed by medium to poor (32%) and rich (31%). The literacy rate across the villages was however quite satisfactory- about 54% with children who read at the primary level comprised the most (61%).

*Households dependency on forests*

About 13% of the households in the sampled villages were highly dependent on forest for their livelihoods, whereas, the remaining HHs were moderately or lastly dependent on forest (Figure 2). The dependency rate, however, varies with their location and with socio-economic condition of the corresponding HH. In Tiprapara dependency rate was significantly hight ($F=2.72$; $p=0.11$) due to its location within the national park. In Tiprapara majority (67%) of the people was mostly depenedent on the park for their livelihoods ($t=5.94$, $p\leq0.01$), whereas 29% of the people from Ratanpur were moderately dpenedent on the park for their livelihoods ($t=3.38$, $p\leq0.05$).

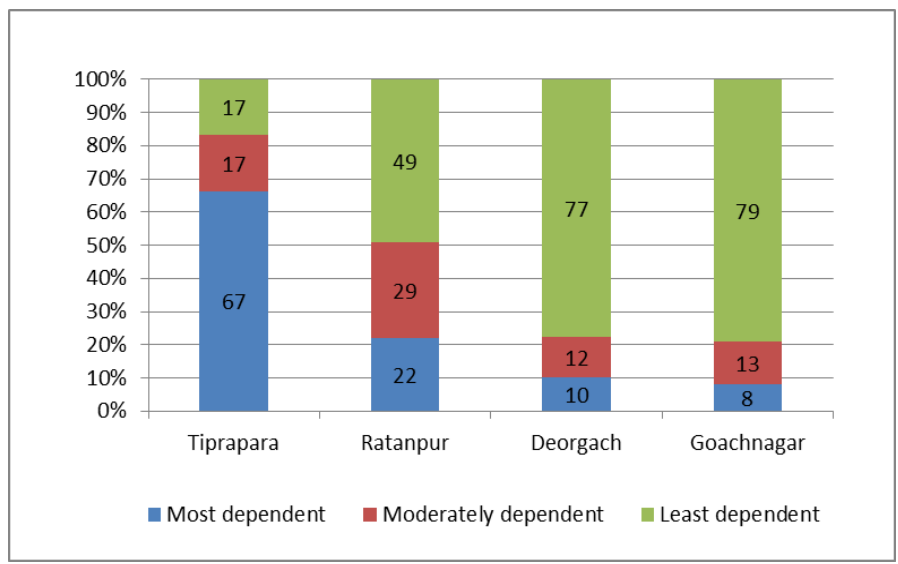

**Figure 2.** Forest dependency across the study villages

Figure 3 shows the relative financial contribution of major livelihood activities among the HHs in Satchari area. Agriculture was the major source of income of HHs in the villages contributing 30% of their annual income followed by small-scale business (21%). Collection and selling of various NTFPs constituted about 19% of HHs annual income followed by timber poaching (illegally) from the RF (11%).. In Tiprapara all the households depends on forest for their domestic biomass fuel requirement for cooking. They also cultivated lemon in a confined area of the PA designated by the park authority.

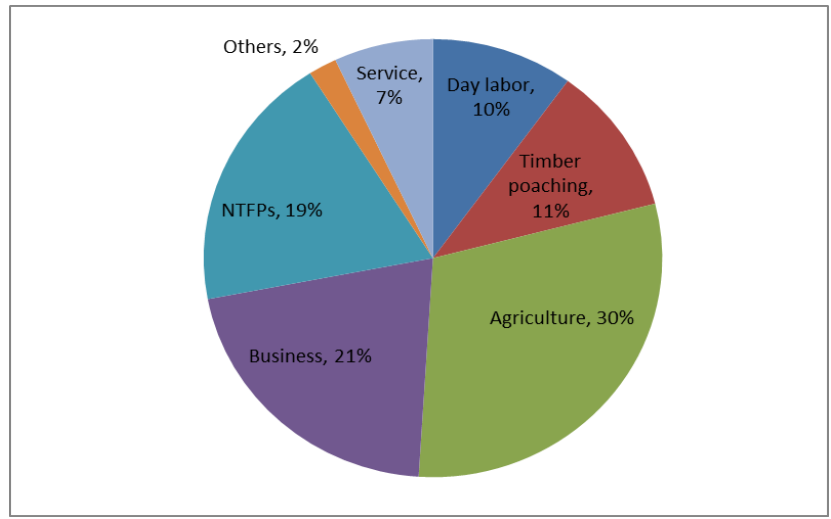

**Figure 3.** Relative financial contribution of different sectors to local livelihoods in Satchari

*The collection and use of NTFPs*
Altogether we found fourteen different NTFPs which households harvested either from the national park (PA) or from the surrounding forests (RF) regularly or occasionally (Table 3). However, only a few of these made a significant contribution to HHs income. Mainly five NTFPs—firewood, medicinal bark of *menda*, *taragota*- a substitute of cardamom, bamboo and *tendu* leaves which is used to wrap tobacco—account for more than 90% of NTFPs-based income in the area. Collections of those NTFPs however, were not identical in all the villages. All the HHs of Tiprapara collected firewood from the area, while 60%, 55% and 56% HHs respectively from Ratanpur, Deorgach and Goachnagar collected firewood from the area (Figure 4). Besides, *taragota* was mostly collected by the villagers of Ratanpur ($t=3.93$, $p=0.29$) and *menda* bark was found to collect mostly by the people of Deorgach ($t=4.14$, $p=0.26$). Medicinal plants possessed a great diversity among NTFPs in the area; approximately 25% households from the study villages reported at least uses of certain medicinal plants for treating common ailments. During the survey we recorded a total of forty medicinal plants with regular uses in the area (see Mukul et al. 2007 for detail).

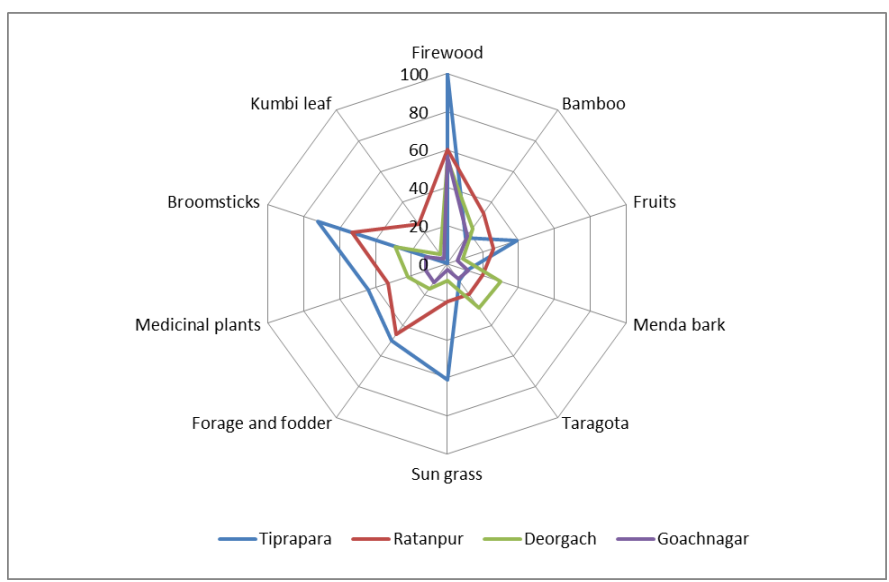

**Figure 4.** Households' involvement (%) in different NTFPs collection in Satchari

Households were found to collect NTFPs both from the PA as well as from the RF (Table 3). About 36% of the respondents collected NTFPs only from the PA, and 41% reported both PA and RF as their NTFPs source where the rest 23% mentioned RF as their sole source of collection. The three major commercial NTFPs - firewood, *taragota* and *kumbi* leaf were mainly collected from the PA. Other NTFPs like, fruits, sungrass, bamboo, *menda* bark and broomsticks though were also collected from the PA, but major proportion were actually came from the surrounding RF.

**Table 3.** Different NTFPs exploited from the SNP and adjacent forests by local communities

| Products | Biological origin | Extent of collection* | Source** |
|---|---|---|---|
| Bamboo | *Bambusa vulgaris* Schard. *Melocanna baccifera* Roxb. | Medium | PA, RF |
| Broomsticks | *Thysanolaena maxima* Roxb. | Medium | PA |
| Bushmeat | - | Very low | PA, RF |
| Firewood | All woody species | High | PA, RF |
| Forage and fodder | - | Low | RF |
| Fruits | *Artocarpus heterophyllus* Lamk. *Artocarpus chaplasha* Roxb. *Artocarpus lakoocha* Roxb. *Citrus limon* L. *Syzygium spp.* | Low | PA, RF |
| Honey | *Apis florae* *Apis dorsata* | Very low | RF, PA |
| Kumbi leaf | *Careya arborea* Roxb. | Medium | PA |
| Medicinal plants | - | Low | PA, RF |

| Menda bark | *Litsea monopetala* (Roxb.) Pers. | Medium | PA, RF |
| Rattan | *Calamus guruba* Ham. | Low | RF, PA |
| | *Daemonorops jenkensianus* Mart. | | |
| Sand | - | Medium | PA, RF |
| Sun grass | *Imperata cylindrica* L. | Medium | PA, RF |
| Taragota | *Ammomum aromaticum* Roxb. | Medium | PA |

\* Based on peoples perception
\*\* PA- protected area, i.e., Satchari National Park; RF- reserved forest, sequence indicates their relative importance as source.

*The importance of NTFPs to local livelihoods*

NTFPs were found critically important as HHs primary, supplementary and emergency source of income. Twenty seven percent of the sample HHs received at least some cash income from the extraction and trading of NTFPs. This contributed on an average 19% of HH's net annual income. Again, collections, processing and sale of NTFPs constituted primary occupation for approximately 18% of the HHs. Study revealed that, HHs belongs to extremely poor category received the most finacial benefits ($t$=1.50, $p$=0.27), and medium to poor HHs moderately dependent on NTFPs for their income ($t$=2.52, $p$=0.12) . The importance of NTFP-based income across the three distinct income groups are shown in Figure 5. The net yearly income from NTFPs per HH was varied bewteen Tk. 2500 to Tk. 25000, (i.e. Tk. 40 to 120 daily).

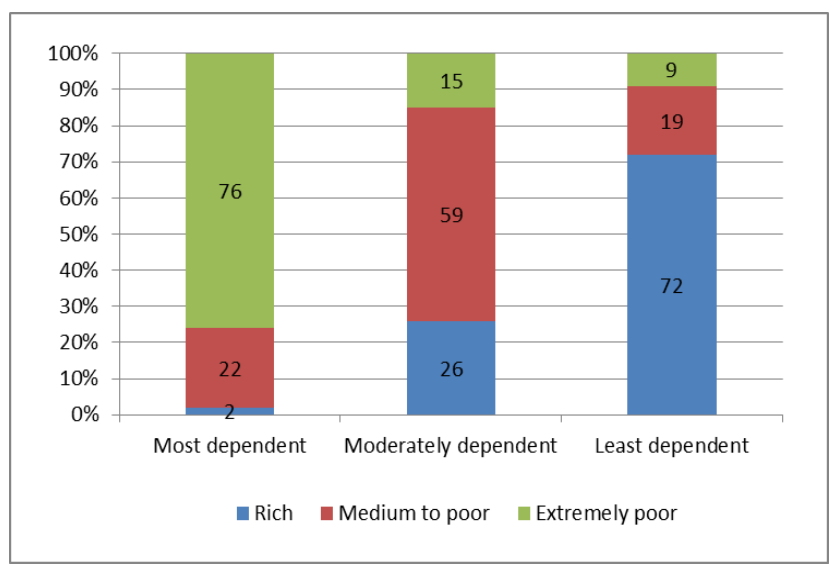

**Figure 5.** Comparison of financial gain from forest products and NTFPs among different income groups

When considering the trade, firewood was the most collected NTFPs in the area, and people directly brought it to the local market to directly sell that to consumers or to retail shops. The others commercially important NTFPs like, dried *taragota* were purchased by the middlemen (intermediary agents), who in their turn sold it to the *Ayurvedic* medicine (a form of herbal medicine) company located in the capital city Dhaka; *menda* bark was directly sold to local *menda* based mosquito repellent factories as raw ingredients. Bamboo, another important NTFP was also sold to local market directly by the collectors themselves. Some other NTFPs such as sungrass, broomsticks, fruits, medicinal plants, rattan and sand were mainly collected for households self consumption and rarely sold to the local markets although they posses a great market value too. Rich HHs owned majority of the NTFPs based enterprise and retail shops in the area and profited mostly (i.e., get maximum returns on their investments) from the trading of NTFPs, although they were not essentially involved in the collection and gathering of NTFPs. This is due to the limited access of rural NTFPs collectors in regional market, poor storage and technical facilities.

*NTFPs as a mean of improving households' resilience during emergencies*
Though dependency of households on NTFPs in the area was limited to 18% and 19% in terms of primary occupation and contribution to their net annual income but NTFPs was found contributing considerably during HHs tough economic situation. Such situation was attributed by agricultural crop failure; reduction in the price of the agricultural crops they produced; increases in the price of essential farm materials like irrigation, fertilizers and labor ; l common naturacatastrophes in the area like flood and drought; monetary inflation limiting households buying capacity, and disease of the family members incurring unplanned cost related to treatment. Table 4 below summarizes the households' opinion on the role of NTFPs during such type of harsh situation as an option to increase their resilience capacity.

**Table 4.** Respondent's responses on the perceived value of NTFPs in households' resilience

| Comment/ Opinion | Number of respondents | | |
|---|---|---|---|
| | Yes | No | No idea/comment |
| NTFPs keep an open option to depend on during the time of emergencies | 49 | 6 | 46 |
| NTFPs contribute directly to hh's adaptation strategies | 36 | 47 | 18 |
| NTFPs contribute indirectly to hh's adaptation strategies | 41 | 36 | 24 |
| NTFPs increases hh's resilience capacity | 61 | 12 | 28 |

**Discussion**
Although the present study covers only one PA of Bangladesh, we believe it fairly represents the overall situation in PAs throughout the country, a country overwhelmed with high population density, resource scarcity, and extreme pressure on its limited forests (Mukul et al. 2010, Sohel et al. 2014). Our study could also potentially represent other parts of south and south-east Asia that possess similar socio-political and economic features. In Satchari, we observed a strong link between poverty and dependence on forest, and forests were more important to low-income than to high-income people. In the area, forest contributed nearly all the aspects of rural life; providing food, fodder, fuel, medicines and building materials, as well as intangible benefits which is difficult to measure in terms of monetary value. Local inhabitants have traditionally used the national park and the RF for sustaining their livelihoods. It was, however, apparent that, dependencies of HHs on national park and RF increasesd with the decrease of their distance from the forest, and HHs with higher socio-economic status depends less on forests. This are in accordance with the findings of Davidar et al. (2010) and Fisher et al. (2010) who also found location of the HHs as an influential factors of HHs forest products collection respectively in India and Malawi. The dependency of poorest people on NTFPs-based income is also evident from the study of Cavendish (2000) and Malhotra et al. (1991) conducted respectively in Zimbabwe and India. The greater significance and dependence on NTFPs of low-income communities (compared to high-income groups) are also apparent from the work of Falconer (1992), Cavendish (1997), Pimentel et al. (1997), Neumann and Hirsch (2000), and Angelsen and Wunder (2003).

Twenty seven percent of the HHs in our study, most of which were poor to extremely poor, received at least some cash benefits from NTFPs, that constituted an estimated 19% of HH's net annual incomes, and NTFPs were a primary source of earnings for about 18% of the HHs. These figures are comparable with the findings of Das (2005), who assessed the role of NTFPs among forest villagers of the Buxa Tiger Reserve (a PA) in India. Das (2005) found that 54% of the households of the reserve were dependent on NTFPs, with 26% of the HHs reported that, NTFPs were their major source of income comprising 40% of their net annual income. Table 5 compares the finding of our study with several earlier studies that also presents the extent of contribution of NTFPs to local livelihoods in and around protected area. In Satchari area, we observed that the poorer people failed to optimize the profits they might have earned from selling NTFPs because of limited access to technology, capital, and market facilities. They were overwhelmingly dependent on intermediaries to dispose of the products they obtained, a phenomenon also mentioned in the studies of Arnold and Ruiz-Perez (2001) and Shackleton and Shackleton (2004).

**Table 5.** The extent of contribution of NTFPs to local people's cash income: some examples

| Country | Estimates of contribution | Author(s) |
| --- | --- | --- |
| Orissa, India | 19% | Mahapatra et al., (2005) |
| Jharkhand, India | 11% | - |
| Cameroon | 15% | Ambrose-Oji (2003) |
| Zimbabwe | 22%-35% | Cavendish (2000) |
| Tamil Nadu, India | 24% | Ganesan (1993) |
| Sri Lanka | 21% | Gunatillike et al., (1993) |
| West Bengal, India | 17% | Malhotra et al., (1991) |

Shackleton and Shackleton (2004), in a study in South Africa, reported the role of NTFPs as a safety net during the time of emergencies. Fisher et al. (2010), in Southern Malawi, found that local communities in response to climate-induced vulnerabilities and food shortages depends on forests as a key coping strategy. They did this by sourcing tubers and forest fruits to meet their dietary requirements, and through sourcing timber and NTFPs to supplement their incomes (Fisher et al. 2010).

**Conclusion**

Our study has reiterated the critical role of NTFPs in providing subsistence and cash incomes to local communities, especially to the poorer groups, and as a provisioning option during unforeseen events that improves HHs' resilience capacity. Recent renewed emphasis on conservation in PAs sometimes lead to restrictions on extraction of NTFPs. Such restrictions need to be viewed and considered within the broader context and reality of the high degree of dependence of poorer communities on NTFPs. A degree of flexibility in existing PA management may be warranted, such as by setting an allowable resource extraction limit for deserving community members in the light of both ecological and economic sustainability. Some species are particularly vulnerable to over-exploitation, and require special attention from PA managers. *L. monopetala,* in our study, for example, has experienced serious depletion due to unsustainable bark collection and rampant illegal removal. Policy makers and park managers should consider improving awareness of local communies about sustainable harvesting of NTFPs. They could also allow them to cultivate commercially important NTFPs in buffer zones to reduce pressure on core PAs. Another important issue is the lack of organized and equitable market outlets and facilities for local forest dependent communities to sell the NTFPs they collected. It makes them entirely dependent on an exploitative network of intermediaries. PA managers could therefore alsofacilitate direct and wider access of the poorer HHs to markets.

**Note**

An abridged version of this paper is available at - Fox, J., B. Bushley, S. Dutt, and S.A. Quazi. 2007. *Making Conservation Work: Linking rural livelihoods and protected areas in Bangladesh,* eds. Hawaii: East-West Centre and Dhaka: Nishorgo Support Project of Bangladesh Forest Department.

**Acknowledgements**

We extend our gratitude to the local communities and participants in the study area (Tiprapara, Ratanpur, Deorgach and Goachnagar). Thanks to Mr. Mashiur R.Tito and Mr. MASA Khan for accompanying us during the field visits. Our thanks also due to Dr. Brian Bushley and Dr. Shimona A. Quazi of University of Hawaii for their invaluable feedback on an earlier version of this work, and USAID - Bangladesh for funding the fieldwork. The valuable feedback from the anonymous reviewers and editor greatly improved the quality of this manuscript.